\documentclass[letterpaper]{epl2}

\usepackage{amsmath}
\usepackage{graphicx}
\usepackage{bm}
\newcommand{\<}{\langle}
\renewcommand{\>}{\rangle}

\title{The Berry phase and the pump flux in stochastic chemical
  kinetics}
\shorttitle{The Berry phase in chemical kinetics}

\author{N.A. Sinitsyn \inst{1,2} \and Ilya Nemenman \inst{2}}
\shortauthor{N.A. Sinitsyn and Ilya Nemenman}

\institute{
  \inst{1}Center for Nonlinear Studies and \\
  \inst{2}Computer, Computational and Statistical Sciences Division,
  Los Alamos National Laboratory, Los Alamos, NM 87545 USA
}

\abstract{ We study a classical two-state stochastic system in a sea
  of substrates and products (absorbing states), which can be
  interpreted as a single Michaelis-Menten catalyzing enzyme or as a
  channel on a cell surface. We introduce a novel general method and
  use it to derive the expression for the full counting statistics of
  transitions among the absorbing states. For the evolution of the
  system under a periodic perturbation of the kinetic rates, the
  latter contains a term with a purely geometrical (the Berry phase)
  interpretation. This term gives rise to a pump current between the
  absorbing states, which is due entirely to the stochastic nature of
  the system. We calculate the first two cumulants of this current,
  and we argue that it is observable experimentally.  }

\begin{document}

\maketitle


Single molecule experiments \cite{english-etal-06} have led to a
resurgence of work on the stochastic version of the classical
Michaelis-Menten (MM) enzymatic mechanism that describes the
enzyme-driven catalytic conversion of a substrate into a product
\cite{MM}. A lot of progress (analytical and numerical) has been made
under various assumptions and for various internal enzyme structures
\cite{qian-elson-02,rao-arkin-03,qian-xie-06,gopich-szabo-06,xue-etal-06}. Still,
many questions linger. Specifically, linear signal transduction
properties of biochemical networks, such as frequency dependent gains,
are of a high interest
\cite{detwiler-etal-00,samoilov-etal-02,samoilov-etal-05}. These are
studied by probing responses to periodic perturbations in the input
signals (in our case, kinetic rates) due to fluctuations in chemical
concentrations, in temperature, or due to other signals.  However,
such approaches usually disregard a phenomena known from the theory of
stochastic ratchets \cite{reimann-02}, especially in the context of
biological transport
\cite{julicher-etal-02,parrondo-02,qian-98,astumian-bio98}. Namely, a
system's symmetry may break, and, under an influence of a periodic or
random zero-mean perturbation, the system may respond with a finite
flux in a preferred direction. This, indeed, happens for the MM
reaction \cite{astumian-03}, and we study the phenomenon in detail in
this work.

The ratchet or pump effect manifests itself during periodic driving of
simple classical systems, such as a channel in a cell membrane, which
is formally equivalent to the MM enzyme
\cite{westerhoff-86,chen-87,astumian-87,tsong-88,astumian-pra89,astumian-jchph89,robertson-90,robertson-91,markin-91,astumian-03,tsong-chang-02}. The
prevalence of the transport terminology over the chemical one in the
literature is grounded in experiments, where driving is achieved by
application of periodic electric field or nonequilibrium noise that
modulate barrier heights for different channel conformations. The
resulting cross-membrane flux has contributions that have no analog in
stationary conditions.
 
In a related system, a quantum pump
\cite{brouwer-98,makhlin-mirlin-01,moskalets-buttiker-02}, pump fluxes
admit a purely geometrical Berry phase interpretation
\cite{pump_berry}. A similar interpretation has been introduced for
some special classical cases \cite{shi,hannay-85}.  However, the Berry
phase has not yet been derived for classical stochastic Markov chains
that model chemical kinetics, even though recent developments strongly
hint at the possibility
\cite{hill-book,qian1-98,qian-00,astumian-03,ao-04,kwon-05}.  For
example, such systems admit introduction of a vector potential for the
fluxes to characterize circular motion \cite{qian1-98}, allow
reformulation of the Langevin dynamics in the form that strongly
resembles wave packet equations in quantum mecanics with a nontrivial
Berry curvature of Bloch bands \cite{ao-04,kwon-05,niu}, and result in
pump currents that depend on contour integrals over the the evolution
of the system's parameters \cite{astumian-03}.  Although currents in
most analyzed models were produced by the lack of the detailed balance
rather than by external time-dependent perturbations, a dual
description in terms of an external noise induced ratchet effect is
usually possible.  Conversely, time-dependent perturbations can break
the detailed balance and induce the catalytic cycle \cite{chen-87}.

In the present work we demonstrate that a theory based on the Berry
phase can be constructed for a purely classical adiabatically slowly
driven stochastic dynamics. The theory leads to an elegant
interpretation of prior results, and it makes new predictions even for
the minimal model of the MM reaction. Additionally, the theory fills
in an important gap in the current state of the field by providing a
simple, yet general recipe for calculation of mean particle fluxes and
their fluctuations.

\begin{figure}[t]
\centerline{\includegraphics[width=6 cm]{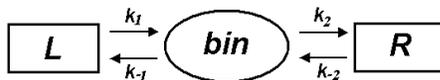}}
\caption{\label{system}Transition rates into and out of the absorbing
  states $L$ (substrate) and $R$ (product). Note that the rates $k_1$
  and $k_{-2}$ denote total rates (rather than per mole of
  substrate/product), and they can be modulated, for example, by
  changing concentrations of the substrate and the product,
  respectively.}
\end{figure}

In a standard minimal model, a single MM enzyme catalytic reaction
consists of an enzyme-substrate complex formation and its decay into
the enzyme and the product.  Both reactions may be reversible
\cite{qian-elson-02}.  We assume no multiple internal enzyme states
(see \cite{english-etal-06,qian-xie-06} for the complementing
discussion). The reaction can be viewed as a single bin (the enzyme)
that can contain either zero or one substrate particles in it. The
particle can escape from the bin by jumping into one of the two
absorbing states, either returning to the substrate -- the Left, or
converting into the product -- the Right, cf.~Fig.~\ref{system}.  In
turn, if the bin is empty, either of the absorbing states can emit a
new particle into it.  All transition times are exponentially
distributed with the rates defined as in Fig.~\ref{system}.  We
investigate the mean particle current $J$ from $L$ into $R$ on time
scales much larger than the typical transition times between any two
states. Our main goal is to understand the effect of periodic driving,
i.\ e., periodic rate changes.

We emphasize again that this system is equivalent to a cross-membrane
transport problem, where the L/R states correspond to compartments on
different sides of the membrane, and the bin is a membrane channel
complex.  Additionally, this system is relevant to transport through a
quantum dot in the Coulomb blockade regime \cite{nazarovFCS}.
 

Let $P_{e}$ and $P_{f}$ be probabilities of an empty/filled bin
(unbound/bound enzyme). Neglecting particle discreteness and assuming
that all rates change slowly, so that $P_{e/f}$ are always in
equilibrium, one derives for the instantaneous $L\to R$ current
\begin{equation}
  j_{\rm cl} (t)=\frac{\kappa_+(t)-\kappa_-(t)}{K(t)},\,\; K\equiv \sum_{m}
  k_m,\,\; \kappa_{\pm}=k_{\pm1}k_{\pm2},
\label{jcl}
\end{equation} 
and the summation is over $m=-2,-1,1,2$.  This is the classical (MM)
current. Notice that, even if $\<k_1k_2\>=\<k_{-1}k_{-2}\>$ (where
$\<\dots\>$ denotes time averaging), $J_{\rm cl}\equiv\<j_{\rm cl}\>\neq0$
due to nonlinearities of the MM reaction.

How is this result affected by the particle discreteness? The master
equation for the system's evolution is:
\begin{equation}
\frac{d}{dt} \left[ \begin{array}{l}
P_{e}\\
P_{f}
\end{array} \right] = - \left[ 
\begin{array}{ll}
\,\,\,\,k_1+k_{-2}   &  -k_{-1}-k_2\\
 -k_1-k_{-2}    &\,\,\,\, k_{-1}+k_2
\end{array} \right]
\left[ \begin{array}{l}
P_{e}\\
P_{f}
\end{array} \right],
\label{ev1}
\end{equation}
and we assume the adiabatic approximation, i.\ e., the rates $k_m$
oscillate with the frequency $\omega\ll \min_m k_m$. Since
$P_e=1-P_f$, (\ref{ev1}) is equivalent to a formally solvable first
order linear differential equation. However, expressing the solution
in a simple form requires approximations (e.g., small fluctuations)
even for a harmonic dynamics of the parameters
\cite{astumian-03}.  Additionally, the solution provides little
information about the fluxes beyond their mean values, and it is not
easy to generalize to the case of more complicated, multi-state
stochastic systems. Thus here we pursue a different analysis
technique.
 
The formal solution of (\ref{ev1}) is
\begin{equation}
{\bf p}(t)=\hat{T}\left(e^{-\int_{t_0}^{t}\hat{H}(t)dt}\right){\bf p}(t_0).
\label{ev12}
\end{equation}
where $\hat{T}$ stands for the time-ordering operator, ${\bf
  p}(t)\equiv [ P_{e}(t),P_{f}(t) ]^T$, and
\begin{equation}
\hat{H}= \left[ 
\begin{array}{ll}
\,\,\,\, k_1 +k_{-2}   & -k_{-1}-k_2 \\
-k_1-k_{-2}   & \,\,\,\, k_{-1}+k_2
\end{array} \right].
\label{h0}
\end{equation}

It is useful to separate the parts of $\hat{H}$ that are directly
responsible for the transition into or out of the $R$-state, namely
$\hat{H}\equiv\hat{H}_0 - \hat{V}_{+}-\hat{V}_{-}$, where 
\begin{eqnarray}
&&\hat{H}_0= \left[
\begin{array}{cc}
k_1+k_{-2}   & -k_{-1} \\
-k_1   &  k_{-1}+k_2
\end{array} \right],
\label{h00}\\
&&\hat{V}_+= \left[ 
\begin{array}{cc}
0   & k_2 \\
0   &   0
\end{array} \right], \,\,\,\,\,\,\,\,
\hat{V}_-= \left[ 
\begin{array}{cc}
0   & 0 \\
k_{-2}   &   0
\end{array} \right].
\label{VV}
\end{eqnarray}
With this, we can write a formal expression for the probability $P_n$ to
have $n$ net transitions from the bin into $R$ during time $t$. This
is similar to \cite{gopich-szabo-06}, but we also allow $k_{-2}\neq0$,
counting transitions from $R$ into the bin with the negative sign. For
example,
\begin{align}
P_{n=1}&= {\bf 1}^+\left[ \int_{t_0}^{ t} dt_1 U_0(t,t_1)  
\hat {V}_{+}(t_1) U_0(t_1,t_0) \right.+ \nonumber\\
+&\int_{t_0}^{t} dt_3 U_0(t,t_3)  
\hat {V}_{+}(t_3)
\int_{t_0}^{t3} dt_2 U_0(t_3,t_2)  
\hat {V}_{-}(t_2)\times \nonumber\\
\times&\left.\int_{t_0}^{t2} dt_1 U_0(t_2,t_1)\hat {V}_{+}(t_1)
  U_0(t_1,t_0) +\cdots\right] {\bf p}(t_0). \label{p1}
\end{align}
Here ${\bf 1}$ is the vector with all unit entries and
\begin{equation}
U_0(t,t') = \hat{T}\left(e^{-\int_{t'}^t\hat{H}_0(t)dt}\right).
\label{u0}
\end{equation}
We introduce the probability generating function
(pgf) for the number of transitions from the bin into R,
\begin{equation}
Z(\chi)=e^{S(\chi)}=
\sum_{s=-\infty}^{\infty} P_{n=s}e^{is\chi}.
\label{pgf1}
\end{equation}
$\chi$ is called {\em the counting field}, $S(\chi)$ is {\em the full
  counting statistics}, and its derivatives give cumulants (or
connected correlation functions) of $P_n$, e.g. $\<n\>=-i \partial
S(\chi)/\partial \chi |_{\chi=0} $, $\<\delta^2
n\>\equiv\<n^2\>-\<n\>^2 = (-i)^2 \partial^2 S(\chi)/\partial \chi^2
|_{\chi=0}$, etc.

We derive the full counting statistics similarly to
\cite{gopich-szabo-06,nazarovFCS}, but we extend the method by
allowing for an explicit time-dependence of $\hat{H}$.  First notice
that (\ref{pgf1}) with the expansion of probabilities as in (\ref{p1})
is equivalent to an expansion of an evolution operator with a
$\chi$-dependent Hamiltonian:
\begin{equation}
Z= {\bf 1}^+\hat{T}\left(e^{-\int_{t_0}^{t}\hat{H}(\chi,t) dt}\right) {\bf p}(t_0),
\label{pdf2}
\end{equation}
where 
\begin{align}
\hat{H}(\chi,t)&=\hat{H}_0(t)-\hat{V}_+(t)e^{i\chi}-
\hat{V}_{-}(t)e^{-i\chi}\nonumber\\
=&\left[ 
\begin{array}{cc}
k_1 + k_{-2}   & -k_{-1}-k_2 e^{i\chi} \\
-k_1-k_{-2}e^{-i\chi}   & k_{-1}+k_2
\end{array} \right].
\label{hchi}
\end{align}
We define the two instantaneous eigenstates of $\hat {H}(\chi,t)$ as
$|u_0\rangle$ and $|u_1\rangle$. There are also the left eigenstates $
\langle u_0|$ and $\langle u_1|$ with the same eigenvalues; we
normalize them so that $ \langle u_{n}(t)|u_m (t)\rangle=\delta_{nm}$.

Let's introduce an intermediate time scale $\delta t$, $ 1/\omega \gg
\delta t \gg \max_m (1 /k_m)$.  During this time, typically, many
transitions happen but $k_m$'s are approximately constant.  Consider
time-points $t_j=t_0+j\delta t$, $j=1\dots N$, and $t_N\equiv t$.  In
the adiabatic limit, one can approximately rewrite the expression for
the time ordered exponent in (\ref{pdf2}) as a product of evolution
operators over time intervals of the size $\delta t$ with parameters
set to constants at times $t_j$, i.\ e.,
\begin{equation}
Z\approx  {\bf 1}^+ e^{-\hat{H}(\chi,t_N)\delta t}
e^{-\hat{H}(\chi,t_{N-1})\delta t} \dots e^{-\hat{H}(\chi,t_0)\delta t} {\bf p}(t_0).
\label{ZZ}
\end{equation}
Now we insert the resolution of the identity
\begin{equation}
\hat{1}= |u_0 (t_i)\rangle \langle u_0(t_i)|
+ |u_1 (t_i)\rangle\langle u_1(t_i)|
\label{one}
\end{equation}
after every exponent at $t_i$. We define $|u_0(t_i) \rangle$ as the
eigenstate of $\hat{H}(\chi,t_i)$ corresponding to
the eigenvalue $\lambda_0$ with the smaller real
part. Since $\delta t\gg1/k_i$, $\vert \exp\left[-\lambda_0(\chi,t_i)\delta
  t\right]\vert \gg \vert \exp\left[-\lambda_1(\chi,t_i)\delta t\right]\vert$, and
terms containing $|u_1(t)\rangle$ can be neglected. Moreover, after
long evolution, any information about the initial and the final states
is lost. Thus we rewrite the pgf as
\begin{equation}
Z\approx e^{-\lambda_{min}(\chi,t_0)\delta t} \prod_{i=1}^Ne^{-\lambda_0(\chi,t_i)\delta t} 
\langle u_0(t_i)|u_0(t_{i-1})\rangle.
\label{ZZ1}
\end{equation}
Finally, we approximate $\langle u_0(t_{i})|u_0(t_{i-1})\rangle
\approx \exp[-\delta t \langle u_0(t_{i})
  | \partial_t|u_0(t_{i-1})\rangle]$, which allows us to write
\begin{equation}
  S(\chi) \approx S_{\rm geom}+S_{\rm cl}=-\frac{T}{T_0}\int_{0}^{T_0}dt\, [
  \langle u_0|\partial_t|u_0\rangle+\lambda_0(\chi,t) ],
\label{fcs2}
\end{equation}
where $T_0=2\pi / \omega$ is the period of the rate oscillations and
$T\gg T_0$ is the total measurement time.  The last term in
(\ref{fcs2}) would be the same even for a time-independent problem,
and it has been discussed before \cite{nazarovFCS}. Diagonalizing
$\hat{H}$, and denoting $e_{\pm\chi}\equiv e^{\pm i\chi}-1$, we get for this term
\begin{equation}
  S_{\rm cl}=\frac{-T}{2T_0} \int_0^{T_0} dt \left[ K-
  \sqrt{K^2+4(\kappa_+e_{+\chi}+\kappa_-e_{-\chi})}  \right].
\label{Scl}
\end{equation}
 
The first term in (\ref{fcs2}) is more interesting. Since it depends
only on the choice of the contour in the ${\bf k}$ space, but not on
the rate of motion along the contour (at least in the adiabatic
approximation), it has a geometric interpretation. We write
\begin{equation}
  S_{\rm geom}=-\frac{T}{T_0}\oint_{{\bf c}} {\bf A} \cdot d{\bf
    k},\;\,
  A_m = \langle u_0({\bf k})|\partial_{k_m}|u_0({\bf k})\rangle,
\label{Sgeom}
\end{equation}
where the integral is over the contour in the 4-dimentional 
parameter space (the ${\bf k}$-space) drawn
during one cycle.  

To estimate the error in our results, we note that (\ref{pdf2}) is
exact, and then assumptions followed. First, we neglected the initial
nonequilibrium relaxation on a time scale of $\sim k_m^{-1}$. Since
the $S(\chi)\sim T$, this introduces an error in $S$ of
$\sim1/(k_{m}T)$, which vanishes for long observation times. Second,
we projected the evolution only to the subspace of states with the
smallest real part of the eigenvalue. The resulting error is
exponentially suppressed in the adiabatic limit by
$\exp(-(\lambda_{1}-\lambda_0)/\omega)$. Third, there is the coarse
graining in (\ref{ZZ}).  To the lowest order, it introduces errors in
$S$ in the form of commutators, such as $[H(\chi,t_i),H(\chi,t_j)]$, $|t_i-t_j|
< \delta t$. Since $\langle u_0(t+\delta
t)|[H(\chi,t_i),H(\chi,t_j)]|u_0(t)\rangle \sim O([\omega \delta t]^2)$ for $t
<t_i,t_j<t+\delta t$, this error is less significant than the
$O(\omega)$ contribution from the geometric term in $S$. Finally, the
error due to the approximation of $\<u_0(t_i)|u_0(t_{i-1})\>$ in
(\ref{ZZ1},~\ref{fcs2}) is of the same order.

In a chemical system, the rates $k_m$ can be changed by many means,
e.g., by varying the system's temperature. However, the simplest
scenario is to couple the substrate and the product to particle baths
and to vary the corresponding chemical potentials for both
species. Since the rates $k_1$ and $k_{-2}$ are proportional to the
particle numbers, they will oscillate as well. Thus in what follows we
assume that $k_1$ and $k_{-2}$ are time dependent, while $k_2$ and
$k_{-1}$ are constants.  Then, using Stokes theorem, we write
\begin{equation}
\oint {\bf A} \cdot d{\bf k} = \iint_{{\bf s_c}} dk_{1}dk_{-2} F_{k_1,k_{-2}},
\label{stokes}
\end{equation}
where the integration is over the surface ${\bf s_c}$ enclosed by the
contour ${\bf c}$, and
\begin{equation}
F_{k_1,k_{-2}}=\frac{\partial A_{{-2}}}{\partial k_{1}}-
\frac{\partial A_{1}}{\partial k_{-2}}.
\label{berry}
\end{equation}
We will call $F_{k_1,k_{-2}}$ the {\em Berry curvature} by analogy
with similar definitions in quantum mechanics.  The advantage of
working with $F$ rather than with the {\em potentials} $A_m$ is that
the Berry curvature is gauge invariant, i.e., it does not depend on an
arbitrary ${\bf k}$-dependent normalization of $|u_0({\bf
  k})\rangle$. It is a truly measurable quantity. In our case,
\begin{equation}
F_{k_1,k_{-2}}=\frac{e_{-\chi}(e^{i\chi}k_2+k_{-1})}{[4\kappa_+e_{+\chi}+
4\kappa_-e_{-\chi}+K^2]^{3/2}}.
\label{berry2}
\end{equation}
Note that, with $\chi=0$, the Berry curvature is zero, so it is the
counting field $\chi$ that introduces a nontrivial topology in the
phase space of the eigenstates of $\hat{H}(\chi,t)$. More generally, a
normal Markovian evolution corresponds to $\chi=0$, where the
normalization of $P$ ensures that $F|_{\chi=0}=0$. This may be the
reason why nobody discussed the Berry phases in the context of Markov
chains.

Knowing the full counting statistics, one can study both the particle
flux and its fluctuations. Importantly, from (\ref{fcs2},
\ref{Sgeom}), all cumulants will have the geometric and the classical
terms. In particular, for the mean $L\to R$ flux per unit time, we get
\begin{equation}
  J=J_{\rm pump}+J_{\rm cl}=\iint_{{\bf  s_c}} d^2k 
  \frac{k_2+k_{-1}}{T_0K^3}+
  \int_0^{T_0} dt\, \frac{j_{\rm cl}(t)}{T_0},
\label{JJ}
\end{equation}
where $j_{\rm cl}$ is the classical current defined in (\ref{jcl}).
Generally, the ratio of geometric and classical terms is $\sim\omega
/k_m \ll 1$.

For the large observation time $T$, the total expected $L\to R$
particle flux is $\<n(T)\>=JT$. Similarly, its fluctuations are given
by $\<\delta^2n(T)\>=J^{(2)}T$, where
\begin{align}
&J^{(2)}=-\frac{1}{T}\left.\frac{\partial^2 \left(  S_{\rm geom} + S_{\rm cl}\right)}{\partial \chi^2}\right|_{\chi=0}=J_{\rm pump}^{(2)}+J_{\rm cl}^{(2)}
,\label{noise}\\
&J^{(2)}_{\rm pump}=\iint_{{\bf  s_c}} d^2k 
 \left[ \frac{k_2-k_{-1}}{T_0K^3}- 
 \frac{12(k_2+k_{-1})(\kappa_+-\kappa_-)}{T_0K^5} \right],
\label{pump_noise}\\
&J^{(2)}_{\rm cl}=\frac{1}{T_0}\int_0^{T_0}dt 
 \left[
\frac{\kappa_++\kappa_-}{K}-\frac{(\kappa_+-\kappa_-)^2}{K^3} \right].
\label{cl-noise}
\end{align}
The ratio of the pump and the classical noises is again $\sim \omega/
k_m$. Importantly, we see that, in general, expectations of the total,
the classical, and the pump currents are not equal to their variances.
Thus none of the currents in the MM problem is Poissonian. This is a
general property of complex, multi-step reactions, which is often
neglected in computational studies of biochemical reaction
networks. Also note that $J^{(2)}_{\rm pump}$ in (\ref{pump_noise}) is
not necessarily positive and can, in fact, {\em decrease} the total
noise.  However, the total noise variance $J^{(2)}$ is strictly positive.

While the pump current in this system has been analyzed,
cf.~\cite{astumian-03}, to our knowledge, an expression for $J$
parameterized by the rates, rather than by internal enzyme parameters
has not been available. Even more importantly, expressions for the
noise (either classical or pump) for variable kinetic rates have not
existed either.

The smallness of the pump effect compared to the classical current
complicates its observation. However, several opportunities exist.
One is the dependence of $J_{\rm pump}$ on the frequency of the
perturbation, while $J_{\rm cl}\neq J_{\rm cl}(\omega)$. The second
possibility is to vary the rates along a contour with $J_{\rm
  cl}=0$. However, even in this case the noise may still be dominated
by the classical contribution.

\begin{figure}[t]
\centerline{{\bf(a)}}
\includegraphics[width=8cm]{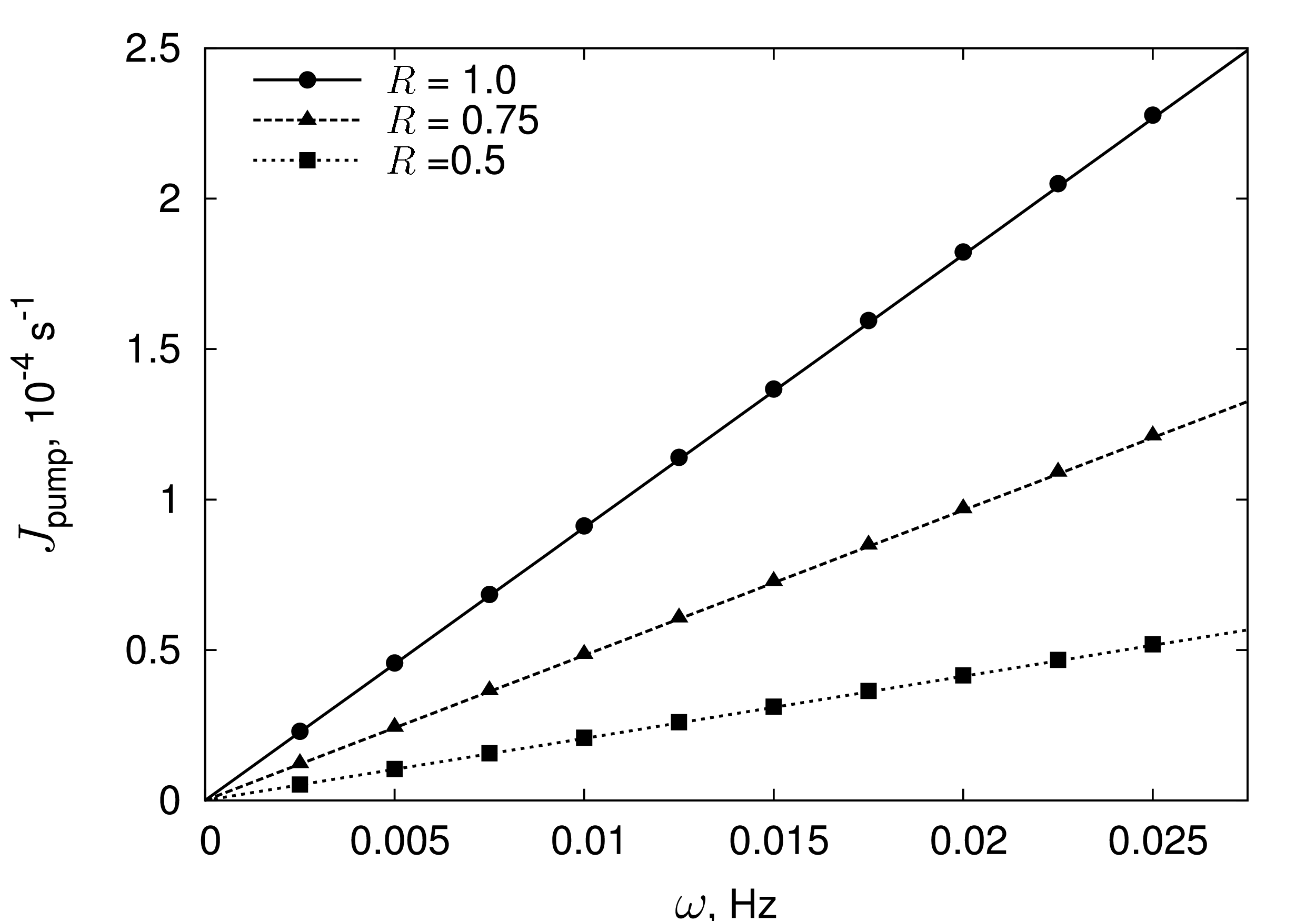} 
\centerline{{\bf(b)}}
\includegraphics[width=8cm]{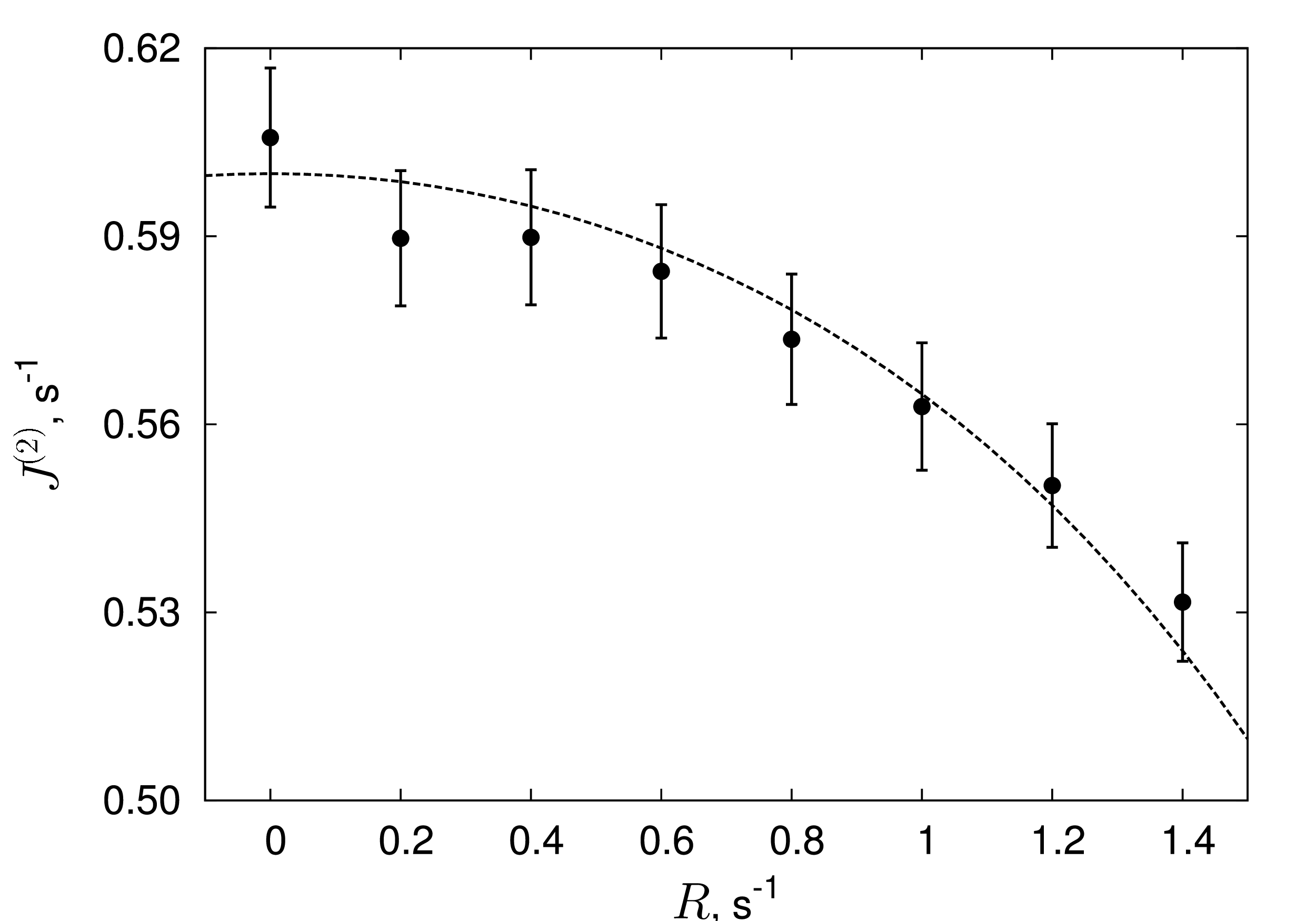} 
\centering
\caption{Comparison of analytical predictions and numerical results
  for a system defined by (\ref{timeev}) with $A=1.5$. Here the theory
  gives: $J_{\rm cl}=0$, $J_{\rm pump}= \omega R^2[2(2-R^2+4A+2A^2)
  ]^{-3/2}$, $J_{\rm cl}^{(2)}=1-(50-2R^2)(25-2R^2)^{-3/2}$, and
  $J^{(2)}_{\rm pump}=0$. Points with error bars mark numerical
  results, while lines correspond to theoretical predictions. {\bf
    (a)} Mean $L\to R$ particle flux $J$ obtained by integrating
  (\ref{ev1}) with $dt=0.001$ and averaged over $\sim 100$ rate
  oscillations. Error bars are negligible, and so is the computational
  time to generate the data.  {\bf (b)} Mean $L\to R$ flux variance
  $J^{(2)}$ for $\omega=0.03$ Hz. Each point is a result of simulating
  $3\cdot 10^6$ elementary particle hops, and covers about 6000 rate
  oscillations. Error bars denote one standard deviation, as estimated
  from the posterior variance of $J^{(2)}$. These simulations took
  less than a day on a modern PC using Octave. }
\label{Jpump}
\end{figure}

To test our predictions, we choose one particular such contour with
$J_{\rm cl}=0$, $J_{\rm pump}\neq 0$ for a numerical analysis:
\begin{equation}
k_{1}=A+R\cos(\omega t),\; k_{-2} =A+R\sin(\omega t),\; k_{-1}=k_2=1.
\label{timeev}
\end{equation}
To estimate $J^{(2)}$ numerically, we implemented a Gillespie-like
scheme \cite{gillespie-77} that admits an explicit time dependence of
the rates. However, since $J_{\rm pump}\ll \sqrt{J^{(2)}_{\rm cl}}$,
such Monte-Carlo estimation of $J$ would take too long.  Instead, we
numerically solved the master equation (\ref{ev1}) for $P_f(t)$ with a
time discretization $dt\ll \min_m (1/k_m)$. This gave a better
precision than an analytical result of \cite{astumian-03}, which
assumes $R\to0$. Knowing $P_f$,
$j(t)=k_2P_f-k_{-2}(1-P_f)$. Fig.~\ref{Jpump} shows an excellent
agreement of our theory with the numerical results for both $J$ and
$J^{(2)}$.


Intuitively, the flux cannot keep increasing linearly as $\omega$
grows and the adiabatic approximation fails.  To understand the
behavior at large frequencies, we consider the case of $\max k_m \ll
\omega$. We look for the probabilities in the form
$P_{e/f}=\bar{P}_{e/f} +\delta P_{e/f}$, where $\bar{P}_{e/f}$ are
calculated for the time-averaged parameter values, namely, $\bar{P}_e
= (\<k_2\>+\<k_{-1}\>)/\<K\>$, $\bar{P}_f=1-\bar{P}_e$, and $\delta
P_{e/f}$ are small and oscillate fast.  The latter can be found from
(\ref{ev1}). For example, when $k_1=\<k_1\> +\delta k_1 \cos(\omega
t)$, $k_{-2}=\<k_{-2}\>+\delta k_{-2} \sin(\omega t)$, and $k_2$,
$k_{-1}$ are constants, we find $\delta P_{e} = -(\delta k_1 \bar{P}_e
/\omega) \sin(\omega t) +(\delta k_{-2} \bar{P}_f/\omega )\cos(\omega
t)$. Then the current averaged over the oscillation period is
$J=J_{\rm cl}(\{\<k_m\>\})+J_{\rm pump}$, where $J_{\rm pump} =
\delta k_1 \delta k_{-2}\bar{P}_e/(2\omega)$ is the correction due to
the fast rate oscillations. For the parameters as in (\ref{timeev}),
we again get $J_{\rm cl}=0$, but $J_{\rm pump}\neq0$. However, the
dependency of $J_{\rm pump}$ on $\omega$ has changed from $\sim\omega$
in the adiabatic case to $\sim 1/\omega$ for fast oscillations. This
agrees with \cite{astumian-03}, which used a different approximation
and showed that there is a single maximum in the pump current at
$\omega \sim k_m$.

The phenomenon of $J_{\rm pump}\neq 0$ can be explained in simple
terms. Since a particle spends a finite time bound to the enzyme, the
values of $k_1$ and $k_{-2}$ cannot influence the system during the
mean unbinding time following a binding event. If, during an
oscillation, the left binding rate is higher than the right one during
the upramp of the cycle, then $k_1$ ``shields'' growing values of
$k_{-2}$ from having an effect, while $k_{-2}$ shields decreasing
values of $k_1$ during the downramp. This leads to a phase-dependent
asymmetry that is the source of the pump flux. Larger frequencies lead
to more shielding, hence the linear $\omega$ dependence. However, for
$\omega\to\infty$, the information about the phase of the oscillations
is lost while a particle is bound, decreasing the asymmetry and the
flux.

Is the pump flux observable experimentally in biochemical (rather than
channel transport) experiments?  Exact zeroing of $J_{\rm cl}$, as in
the numerical example above, which would make $J_{\rm pump}$ the
leading effect, may be difficult to achieve. However, the classical
current is also small near the classical steady state,
$\<k\>_1k_2=k_{-1}\<k_{-2}\>$. In this case, if
$k_{1,-2}=\<k_{1,-2}\>+\delta k_{1,-2}\cos(\omega t-\phi_{1,-2})$, and
if the oscillations are small, $\delta k_{1,-2}/k_{1,-2}\ll 1$, then
we can disregard variation of the Berry curvature inside the contour, and
\begin{align}
J_{\rm pump}&\approx  \omega \frac{k_2+k_{-1}}{2\<K\>^3 }\delta k_1 \delta
k_{-2}\sin\phi,\\
J_{\rm pump}^{(2)}&
\approx  \omega \frac{k_2-k_{-1}}{2\<K\>^3 }\delta k_1 \delta
k_{-2}\sin\phi,
\end{align}
where $\phi=\phi_{-2}-\phi_1$.  This should be compared to the
classical contributions in the same limit
\begin{align}
&J_{\rm cl} \approx \frac{k_2 \delta k_1^2-  
k_{-1} \delta k_{-2}^2 + (k_2-k_{-1})
\delta k_1\delta k_{-2} \cos \phi}{2\<K\>^2},\\
&J_{\rm cl}^{(2)} \approx \frac{2\<k_1\>k_2}{\<K\>}.
\end{align}
We see that, while $J_{\rm pump}/J_{\rm cl} \sim \omega /\<K\>$, the
ratio of the variances is further suppressed, $J^{(2)}_{\rm
  pump}/J_{\rm cl}^{(2)}\sim (\omega/\<K\>)\delta k_1\delta
k_{-2}/(k_1k_{-2})$. Thus overcoming the classical variance is the
biggest concern for a successful experiment.

Our model has a single substrate and a single product. Thus the enzyme
we are discussing is an EC 5 enzyme. Properties of such enzymes vary
dramatically depending on a reaction, a biological species, and
mutations in protein sequences \cite{brenda}. A typical range of $k_2$
{\em in vitro} is $10^{-2}\dots10^4$ s$^{-1}$. Similarly, the
Michaelis constant, $K_M=(k_{-1}+k_2)[S]/k_1$ in our notation, varies
between $0.01$ and $10$ mM (here $[S]$ is the substrate
concentration). For our analysis, we take $k_2\sim 10$ s$^{-1}$, and
$K_M=1$ mM. While little is known about $k_{-1}$ and $k_2$ separately,
it's reasonable to assume $k_{-1}\sim k_2$. Similarly, we take
$k_{-2}\sim k_{1}[P]/[S]$ since both rates are often dominated by the
mean particle-enzyme collision time \cite{qian-elson-02} (for example,
for triose phosphate isomerase reaction, $k_{-2}/k_1\approx 2 [P]/[S]$
\cite{knowles-albery-77}).  Many enzymes with similar parameters have
been characterized in the BRENDA database \cite{brenda}.  Then, with
$[S]\sim [P]\sim 1$ mM and with $\delta k_{1,-2}/k_{1,-2}\sim10$\%, we get
$J_{\rm pump}\sim J^{(2)}_{\rm pump}\sim 10^{-2}/T_0$, $J_{\rm cl}\sim
10^{-1}$, and $J_{\rm cl}^{(2)}\sim 10$ particles per
second. Oscillation periods of $\sim 1$ s are attainable (and still
satisfy $\omega\ll k_m$), which puts the flux ratio at $\sim
10^{-1}$. Different dependence on $\phi$ can further improve
detectability of $J_{\rm pump}$ compared to $J_{\rm cl}$. At these
parameters, the pump flux becomes equal to the total flux standard
deviation for an experiment lasting a few days. Alternatively, working
with, say, $\sim10^6$ enzymes, $J_{\rm pump}$ and $J^{(2)}_{\rm pump}$
become observable in less than a minute. While real experiments will
certainly have additional complications, it is clear that our
predictions should be experimentally testable, at least in principle.

In conclusion, we constructed the Berry phase theory of a purely
classical adiabatic stochastic pump in a simple Michaelis-Menten
enzymatic mechanism. Our approach allowed calculation of the particle
flux and its variance, including the classical and the pump effect
contributions.  We believe that these predictions can be checked
experimentally, and it should be interesting to consider their
importance in the context of enzymatic signal processing.

While we analyzed only one specific model system, the Berry phase
approach is general and can be employed for many processes that can be
reduced to slowly driven Markov dynamics.  Examples range from charge
transport in quantum dots at strong decoherence, to particle transport
through cell membrane channels, and to various biochemical reaction
systems. For these problems, our method has several advantages
compared to already known techniques.  First, it provides a formal
recipe for an analysis of an arbitrary Markov chain: one should
construct a Hamiltonian with counting fields, find its eigenvalue with
the lowest real part and the Berry curvature from the corresponding
eigenstate, and then the counting statistics is given by (\ref{fcs2},
\ref{Sgeom}). Second, the method provides a solution not only for
average fluxes, but also for the full counting statistics and thus for
all flux cumulants. This is important in the context of elimination of
fast degrees of freedom for construction of coarse-grained biochemical
reaction models (e.~g., MM or Hill phenomenological laws), where a
correct treatment of noise in the remaining degrees of freedom has
always been a point of contention. Third, the method allows to
transfer many results of the well-developed Berry phase theory for
dissipative quantum dynamics \cite{garrison-88,vertex-1,vertex-2} to
the field of chemical kinetics. In particular, techniques exist to
extend our work and to calculate flux cumulants to all orders in
$\omega/ k_{min}$ \cite{Garanin}.

\begin{acknowledgments}
  We thank W.\ Hlavacek, F.\ Mu, M.\ Wall, C.\ Unkefer and F.\
  Alexander for useful discussions and critical reading of the
  manuscript. We are grateful to our two anonymous referees for
  insightful comments, which substantially improved this letter. Our
  work was funded in part by DOE under Contract No.\
  DE-AC52-06NA25396.  IN was further supported by NSF Grant No.\
  ECS-0425850.
\end{acknowledgments}

\end{document}